\documentclass[eqsecnum,preprintnumbers,showpacs,showkeys,aps, 10pt]{revtex4}
\usepackage{amsmath, amssymb, bm}

\newcommand*{\ket}[1]{\ensuremath{|#1\rangle}}

\newcommand*{\abs}[1]{\ensuremath{\vert #1\vert}}

\newcommand*{\p}{\ensuremath{\textbf p}}
\newcommand*{\vtau}{\ensuremath{\bm\tau}}

\newcommand*{\bra}[1]{\ensuremath{\langle#1|}}
\newcommand*{\I}{\ensuremath{\mathbb I}}
\newcommand*{\bbK}{\ensuremath{\mathbb K}}
\newcommand*{\bbJ}{\ensuremath{\mathbb J}}
\newcommand*{\bbM}{\ensuremath{\mathbb M}}
\newcommand*{\bbN}{\ensuremath{\mathbb N}}

\newcommand*{\tsigma}{\ensuremath{\tilde\sigma}}
\newcommand*{\hp}{\ensuremath{\hat p}}
\newcommand*{\np}{\ensuremath{\abs\p}}
\newcommand*{\ihalf}{\ensuremath{\frac{i}{2}}}
\newcommand*{\half}{\ensuremath{\frac{1}{2}}}
\newcommand*{\hx}{\frac{X}{2}}
\newcommand*{\hy}{\frac{Y}{2}}
\newcommand*{\X}{\textbf{X}}
\newcommand*{\Y}{\textbf{Y}}
\newcommand*{\Tr}{\text{Tr}}

\begin{document}
\title{Wigner Rotations, Bell States, and Lorentz Invariance \\
of Entanglement and von Neumann Entropy}
\author{Chopin Soo}
\email{cpsoo@mail.ncku.edu.tw}
\author{Cyrus C. Y. Lin}
\email{l2891112@mail.ncku.edu.tw}
\affiliation{Department of Physics, National Cheng Kung University \\
  Tainan 70101, Taiwan.}
\begin{abstract}
We compute, for massive particles, the explicit Wigner rotations of
one-particle states for arbitrary Lorentz transformations; and the explicit
Hermitian generators of the infinite-dimensional unitary representation. For a
pair of spin 1/2 particles, Einstein-Podolsky-Rosen-Bell entangled states and
their behaviour under the Lorentz group are analysed in the context of quantum
field theory. Group theoretical considerations suggest a convenient definition
of the Bell states which is slightly different from the conventional
assignment. The behaviour of Bell states under arbitrary Lorentz
transformations can then be described succinctly. Reduced density matrices
applicable to systems of identical particles are defined through Yang's
prescription. The von Neumann entropy of each of the reduced density matrix is
Lorentz invariant; and its relevance as a measure of entanglement is
discussed, and illustrated with an explicit example. A regularization of the
entropy in terms of generalized zeta functions is also suggested.

\end{abstract}
 \keywords{Wigner rotations, EPR-Bell
states, Lorentz invariance, Entanglement, von Neumann entropy.\\
Published: {\bf Int. J. Quantum Info. {\bf 2} (2004) 183-200}}
\maketitle

\section{Introduction and Overview}

It can be argued that, aside from theories with infinite number of particle
types such as string theory, quantum {\it field} theory is the only way to
reconcile the principles of quantum mechanics with those of special
relativity\cite{weibqft}. In quantum field theory, vanishing correlations for
space-like separated operators are ensured\cite{weibqft}; whereas most efforts
in quantum computation have so far relied upon non-relativistic quantum
mechanics which is not fully compatible with Lorentz invariance and the causal
structure of space-time. Recently however, several groups(see, for instance,
Refs.\cite{milburn,peres,peres2,adami,ahn,terashima,solano,czachorm,milburn2,
terashima2,molotkov,rembielinski}) have focused their investigations on
relativistic effects in quantum information science. The issues include
Lorentz invariance of entanglement, the behaviour of
Einstein-Podolsky-Rosen-Bell states in different inertial frames, and possible
modifications to the degree of Bell inequality violations for moving
observers\cite{czachor}. These relativistic effects may alter the efficiency
of eavesdropper detection in quantum cryptography\cite{ekert3} and compromise
the security of quantum protocols. It is also expected that future applications
in quantum teleportation\cite{bennett4}, entanglement-enhanced
communication\cite{bennett}, high-precision quantum clock synchronization
based on shared entanglement, and quantum-enhanced positioning\cite{josza} will
also require relativistic treatments of quantum systems; and in particular,
the careful analysis of the properties of entangled particles under Lorentz
transformations and the construction of meaningful measures of entanglement.
In this article, we study the behaviour of Bell states under Lorentz
transformations and consider von Neumann entropy as a Lorentz invariant
characterization of entanglement.

We compute, for massive particles, the explicit Wigner rotations of
one-particle states; and the explicit generators of the unitary representation
of the Lorentz group. Unitary representations of the Poincare group acting on
physical states are founded upon Wigner's seminal work\cite{wigner}; but the
explicit expressions of Wigner rotations for massive particles have been
computed, with some difficulty by direct matrix multiplication, only for
rotations and boosts considered separately\cite{ohnuki}. Our derivation,
carried out in Section II, is somewhat simpler, and it permits the explicit
general result for arbitrary infinitesimal Lorentz transformations to be stated
as in Eq.(\ref{e:wrg}). Moreover, with the infinitesimal Wigner angle at hand,
the explicit infinite-dimensional Hermitian generators of the unitary
representation can be worked out, as in Section III. Wigner rotation for a
finite general Lorentz transformation is slightly more complicated, but the
explicit form is also listed in Appendix A.

As basic entangled states, Bell states figure prominently in the literature on
quantum information science and in quantum computational schemes. Their
behaviour under Lorentz transformations is therefore of interest and
importance. In this article we focus on Bell states of two spin $\frac{1}{2}$
massive fermions. Under Lorentz transformations, each one-particle state of
the entangled pair undergoes an $SU(2)$ Wigner rotation. Since
$[SU(2)\times{SU(2)}]/Z_2$ is isomorphic to $SO(4)$, which contains
3-dimensional rotations $SO(3)$ as a sub-group; in group theoretical terms the
entangled system transforms as ${\bf 2}\otimes{\bf 2} = {\bf 4} = {\bf
1}\oplus{\bf 3}$, in which the final step denotes the behaviour under $SO(3)$.
Thus by forming a 4-vector of $SO(4)$ out of the rotational singlet and
triplet Bell states, the complete, and explicit behaviour given the Wigner
angle, of these Bell states under Lorentz transformations can be succinctly
stated, as in Eq.(\ref{e:tyt}). We therefore advocate for Bell states the
convention in Eq.(\ref{e:con}). Details of the setup are presented in Section
IV. In Section V, we introduce reduced density matrices, defined through
Yang's prescription\cite{yang}, for systems of identical particles; and analyze
their behaviour under Lorentz transformations. Without taking Lorentz symmetry
into account, quantum correlations and entanglement of identical fermions have
also been studied in Refs.\cite{ekert,schliemann,yushi,paskauskas}. Here we
are able to show that the von Neumann entropy of each of the constructed
reduced density matrix is Lorentz invariant; and we illustrate the usefulness
of the von Neumann entropy as a Lorentz-invariant measure of entanglement with
a worked example comparing unentangled and Bell states. In the ensuing
subsection, we present a relation between generalized zeta functions and von
Neumann entropy which may be useful for the regularization of the entropy of
infinite-dimensional density matrices. Further comments and conclusions are
presented in the final section.

As far as it is convenient to do so, we shall follow the conventions and
normalizations in Weinberg's tome\cite{weibqft}, with
$\eta^{\mu\nu}=diag(-1,+1,+1,+1)$. Space-time Lorentz indices are denoted by
Greek letters, spatial indices by Latin letters; and summation over repeated
discrete index is assumed. Our computations shall concentrate, for convenience,
only on the homogeneous Lorentz group since translations can be incorporated
rather readily\cite{weibqft} once the behaviour of fields under the homogeneous
group has been worked out.

\section{Wigner rotations}

In quantum field theory, one-particle states $\ket{p,s}$ are classified by
eigenvalues of the Casimir invariants of the Poincar\'e group. Any value of the
momentum, $p^\mu$ can be reached by Lorentz transformation $L(p)$ on a
standard $k^\mu$ for which $p^\mu = L^\mu_{\,\nu}(p) k^\nu$; and the states
can be defined as
\begin{equation}\label{e:sp}
\ket{p,s} \equiv \sqrt{\frac{k^0}{p^0}}U(L(p))\ket{{k},s},
\end{equation}
with $P^\mu\ket{p,s} = p^\mu\ket{p,s}$. It follows(see, for instance,
Ref.\cite{weibqft}) that the effect of an arbitrary Lorentz transformation
$\Lambda$ unitarily implemented as $U(\Lambda)$ on one-particle states is
\begin{equation}\label{e:psfin}
  \ket{p,s}' = U(\Lambda)\ket{p,s} = \sqrt{\frac{(\Lambda p)^0}{p^0}}D_{s' s}
  (W(\Lambda, p))\ket{\Lambda p,s'},
\end{equation}
and
\begin{equation} \label{e:wiggen}
  W(\Lambda,p)=L^{-1}(\Lambda p)\Lambda L(p)
\end{equation}
is a Wigner transformation which leaves $k^\mu$ invariant\cite{wigner}, and
$D(W)$ represents its action on the state (summation convention over the
repeated index ${s'}$ is assumed). The explicit form of $L(p)$ is dependent on
the class of the four-momenta. For massive particles, $p^\mu p_\mu = -m^2 <
0$, and a convenient choice for the standard vector is $k^\mu = (m,{\bm 0})$.
It is then obvious that the set of Wigner transformations leaving $k^\mu$
unchanged is just the rotation group $SO(3)$. Furthermore, $L(p)$ can then be
taken as the pure Lorentz boost
\begin{eqnarray}\label{e:L}
  L^0_{~0}(p)&=&\cosh\chi \\
  L^0_{~i}(p)&=&L^i_{~0}(p)=\hp_i\sinh\chi \\
  L^i_{~j}(p)&=&\delta^i_{~j}+(\cosh\chi-1)\hp^i\hp_j.
\end{eqnarray}
with $\tanh\chi = \frac{\np}{\sqrt{\np^2 + m^2}}$. In this parametrization,
$L(p) =\exp(-i\chi{\hat{\textbf p}}\cdot{\bf K})$ where $K^{i} = M^{0i}$ is the
boost generator\cite{weibqft}.

We may rely upon the analytic nature of Lie groups for the computation of the
Wigner angle, and the corresponding infinitesimal Wigner rotation can be
evaluated by Taylor expansion. Details of the rest of the computations are
delegated to Appendix A. The end result is that the infinitesimal Wigner
rotation of a massive particle is
\begin{eqnarray}\label{e:wrg}
  W(\Lambda,p) &=& \I + \ihalf\left[ \omega_{ij}-\frac{1}{p^0+m}(p_i\omega_{j0}-p_j\omega_{i0})\right]
  M^{ij}\\
  &=& \I + i{\bm\theta}_W\cdot{\bm J},
\end{eqnarray}
with the infinitesimal Wigner angle denoted as
\begin{equation}\label{w1}
  \bm\theta_W  = \bm\theta-\frac{\p \times
  \vtau}{p^0+m} \equiv \bm\theta + {\bm\phi}_1.
   \end{equation}
This agrees with the results of Ref.\cite{ohnuki} when rotations and boosts are
considered separately. It should be noted that the generators of the Wigner
{\it rotations} are indeed $J_i = \frac{1}{2}\epsilon_{ijk}M^{jk}$, but the
complete Wigner angle receives contributions from both the boost and rotation
parameters, $\tau^i= \omega^i\,_0$ and $\theta_i =
\frac{1}{2}\epsilon_{ijk}\omega^{jk}$ respectively, of $\Lambda(\omega)$. In
the absence of boosts, Wigner rotations are degenerate with ordinary rotations
i.e. $\bm\theta_W =\bm\theta$. Although arbitrary finite Lorentz
transformations of $\Lambda(\omega) =
\exp(\ihalf{\omega_{\alpha\beta}}M^{\alpha\beta})$ can be evaluated as
$\lim_{N \rightarrow \infty} [\I +
\frac{i}{2}\frac{\omega_{\alpha\beta}}{N}M^{\alpha\beta}]^{N}$, it is not
simple to express the corresponding finite Wigner rotations in closed form via
products of infinitesimal rotations, essentially because Wigner rotations are
also functions of the momenta. We may however consider a general Lorentz
transformation relating two frames as a product of a pure boost in an
arbitrary direction, $L({\bm\alpha})= \exp(-i{\bm\alpha}\cdot{\bf K})$,
followed by an arbitrary rotation $R({\bm\psi})$. Using the multiplication
rule for Wigner transformations, (Eq.(\ref{e:multi}) in Appendix A), the
closed form Wigner rotation can be expressed as in Eqs.(A16)-(A17). Note also
that for the special case of $U(\Lambda) = U(L(p))$ acting on $\ket{k,s}$ the
Wigner transformation is $W(L(p), k) = [L^{-1}(L(p)k)]L(p)L(k) = L^{-1}(p)L(p)I
= I$, which consistently produces no rotation in spin space, as Eq.(\ref{e:sp})
demands.




\section{Explicit generators of the unitary representation}

One-particle states are defined through the action of creation operator
$a^\dag(\p,s)$ on the vacuum as $\ket{\p,s} = a^\dag(\p,s)\ket{0}$. It follows
from Eq.(\ref{e:psfin}) that under a Lorentz transformation annihilation and
creation operators in quantum field theory behave as
\begin{eqnarray} \label{e:adrule}
 {{a'}^\dag}(\p,s) &=& U(\Lambda){a^\dag}(\p,s){U^\dag}(\Lambda) \\
&\equiv&\int d^3\p \,\,[{{U}^*}_{(\p,s)(\p',s')}(\Lambda)]{a^\dag}(\p',s')\\
&=& \sqrt{\frac{(\Lambda p)^0}{p^0}}D_{s' s}\bigl( W(\Lambda,p) \bigr)
a^\dag(\p_\Lambda, s'),
\end{eqnarray}
with $p^i_\Lambda \equiv \Lambda^i\,_\mu{p^\mu}$; and by taking the adjoint,
\begin{eqnarray}\label{e:arule}
 a'(\p,s) &=& U(\Lambda)a(\p,s)U^\dag(\Lambda) \\
 &\equiv& \int d^3\p \,\,[{U}_{(\p,s)(\p',s')}(\Lambda)]a(\p,s') \\
 &=& \sqrt{\frac{(\Lambda p)^0}{p^0}}D_{s s'}\bigl( W^{-1}(\Lambda,p)
 \bigr)a(\p_\Lambda,s').\label{e:43}
\end{eqnarray}
We may hence deduce from the last equation that
\begin{equation}
{U}_{(\p,s)(\p',s')}(\Lambda) = \sqrt{\frac{(\Lambda p)^0}{p^0}}D_{s s'}\bigl(
W^{-1}(\Lambda,p)\bigr)\delta(\p'-\p_\Lambda),
\end{equation}
and verify by direct calculation, using $D^\dag(W)= [D(W)]^{-1}= D(W^{-1})$,
that the transformation $U(\Lambda)$ is indeed unitary i.e.
\begin{equation}\label{e:45}
\int
d^3\p''\,\,{U}_{(\p,s)(\p'',s'')}(\Lambda){U}^\dag_{(\p'',s'')(\p',s')}(\Lambda)
= \delta(\p-\p')\delta_{s s'}.
\end{equation}
But with the formula of the infinitesimal Wigner rotation of Eq.(\ref{e:wrg})
at hand, it is possible to proceed even further, to obtain the explicit
generators. By considering an infinitesimal transformation with $U(\Lambda) =
\I + \ihalf\omega_{\mu\nu}{\bbM}^{\mu\nu}$ we may express Eq. (\ref{e:arule})
as
\begin{eqnarray}
 &&U(\Lambda)a(\p,s)U^\dag(\Lambda)\\
 &=& a(\p, s) + \ihalf\omega_{\mu\nu}[{\bbM}^{\mu\nu}, a(\p, s)]\\
&=& (\I + \omega^0\,_i\frac{p^i}{2p^0})\biggl(\I -i\bigl[\theta^i +
\omega_{0}\,^k\frac{\epsilon_{ijk}{p^j}}{p^0 + m}\bigr]J^i\biggr)_{s s'}(\I +
\omega^i_\mu p^\mu\frac{\partial}{\partial p^i})a(\p, s') \label{e:44}\\
   &=& \biggl[ \I -i\omega_{0i} \biggl( \underbrace{
     \frac{\epsilon_{ijk}J^j p^k}{p^0+m} + p^0\frac{\partial}{i\partial p^i}
     + \frac{p^i}{2ip^0} }_{\bbK^i}  \biggl)
     - \ihalf \omega_{ij} \epsilon^{ijk} \biggr( \underbrace{ J^k
     + \epsilon^{klm}p^l \frac{\partial}{i\partial p^m} }_{\bbJ^k} \biggr)
       \biggr]_{s s'}a(\p,s') \\
  \label{e:ainvlo}
  &=&  a(\p,s) - \ihalf\omega_{\mu\nu}\left(\int d^3\,\p'\,\bbM^{\mu\nu}_{(\p,s)(\p',s')}
  a(\p', s')\right)
\end{eqnarray}
In terms of creation and annihilation operators, the explicit {\it
infinite-dimensional Hermitian} generators of {\it unitary} transformations
$U(\Lambda) = \exp(\ihalf\omega_{\mu\nu}{\bbM}^{\mu\nu})$ for the {\it
non-compact} Lorentz group are
\begin{eqnarray}
\epsilon^{ijk}{\bbM}_{jk} &=& \int d^3\p \,a^\dag(\p,s)\biggr( J^i
+\epsilon_{ijk}p^j\frac{\partial}{i\partial{p^k}}\biggr)_{ss'}a(\p,s'),\\
{\bbM}^{0i} &=& \int d^3\p\,
a^\dag(\p,s)\biggr(\frac{\epsilon^{ijk}{J_j}p_k}{p^0+m} +
p^0\frac{\partial}{i\partial {p^i}} +
  \frac{p^i}{2ip^0}\biggr)_{ss'}a(\p,s').
\end{eqnarray}
In general, we may also introduce the particle species label $n_i$ for the
creation $a^\dagger(\p,s,n_i)$ and annihilation $a(\p,s,n_i)$ operators; and
the expression of the generators will then include summing over all $n_i$.

It can be verified that the explicit generators
\begin{eqnarray}
  \vec{\bbJ} &=& {\bm J} +\p\times\frac{\partial}{i\partial\p} \\
  \vec{\bbK} &=& (\frac{{\bm
J}\times\p}{p^0+m} + p^0\frac{\partial}{i\partial\p} +
  \frac{\p}{2ip^0}),\label{e:46}
\end{eqnarray}
do satisfy the commutation relations of the Lie algebra of the Lorentz group:
\begin{equation}\label{e:loco}
    [{\bbJ}^i, {\bbJ}^j]= i\epsilon^{ijk}{\bbJ}^k, \qquad [{\bbJ}^i, {\bbK}^j]=
    i\epsilon^{ijk}{\bbK}^k, \qquad
    [{\bbK}^i, {\bbK}^j]= -i\epsilon^{ijk}{\bbJ}^k.
\end{equation}
Likewise $\bbM^{\mu\nu}$ of Eqs.(3.14) and (3.15) obey the similar commutation
relations. The expression of the boost generator of Ref.\cite{czachor}
(following Ref.\cite{ohnuki}) and Ref.\cite{terno} which differs from ours in
not having the final term of Eq.(\ref{e:46}) can be rendered Hermitian provided
the measure $d^3{\p}/p^0$ is adopted instead of Weinberg's.

%
%
%
\section{Bell States and Lorentz transformations}

In this section we specialize to the case with ${\bm J} = {\bm\sigma}/{2}$ for
spin $\frac{1}{2}$ particles. For each massive spin $\frac{1}{2}$ particle, the
Lie group of all Wigner rotations is $SU(2)$. A two-particle state, and in
particular an entangled Bell pair, should transform according to the
$SU(2)\times{SU(2)}$ representation. As we shall see, the isomorphism between
$[SU(2)\times{SU(2)}]/{Z_2}$ and $SO(4)$ permits us to describe the behaviour
of the four basis Bell states of spin space for fixed momenta under Lorentz
transformations succinctly. Since $SO(3)$ is a subgroup of $SO(4)$, in group
theoretical terms if we denote the spin $\frac{1}{2}$ doublet as the ${\bf 2}$
of $SU(2)$, then a 2-particle state behaves as
\begin{equation}
{\bf 2}\otimes{\bf 2} = {\bf 4} = {\bf 1}\oplus{\bf3} \,
\end{equation}
where the final step denotes its behaviour under $SO(3)$. In other words, we
may express the two-particle state in terms of a four-vector (the $\bf 4$) of
$SO(4)$ which transforms as singlet and triplet states under $SO(3)$. Indeed
it is known the four Bell states are expressible as a singlet and a triplet
under ordinary 3-dimensional $SO(3)$ rotations, and as we shall show, they
undergo $SO(4)$ Wigner rotations among themselves under Lorentz
transformations.

The quadruplet $(\mu =0,1,2,3)$ of Bell states can be conveniently defined in
the following manner:
\begin{equation}\label{e:con}
  \ket{B^\mu(\p_1,\p_2)} \equiv \frac{1}{\sqrt{2}}(\tsigma^\mu\sigma^2)_{s s'}
  a^\dag(\p_1,s; n_1)a^\dag(\p_2,s';n_2)\ket{0} \end{equation}
with $\tsigma^0 = i\I_2$ and $\tsigma^i (i=1,2,3)$ being the Pauli matrices
$\sigma^i$. In discussing an entangled pair of identical as well as
distinguishable particles, the additional species labels, $n_{1,2}$, can be
introduced for generality. Let us proceed to show that the states defined
above are indeed Bell states.
The spin indices $s, s'$ are summed over $\pm\frac{1}{2}$ (which we shall
denote as $\pm$ for simplicity). Focusing on the spin part of the states, (and
ignoring for the moment normalization factors and species labels), it can be
checked that the quadruplet is simply\cite{wootters}
\begin{equation}
  \begin{split}
  \ket{B^0}&\propto~~~~\ket{+, 1; -,2} - \ket{-,1; +, 2}  \\
  \ket{B^1}&\propto~~i(\ket{+, 1; +,2} - \ket{-,1; -, 2})  \\
  \ket{B^2}&\propto~~~~\ket{+, 1; +,2} + \ket{-,1; -, 2}  \\
  \ket{B^3}&\propto -i(\ket{+, 1; -,2} + \ket{-,1; +, 2}).
  \end{split}
\end{equation}
For ease of comparison we have also simplified the momenta indices $\p_{1,2}$
to $1,2$ respectively. Therefore, apart from multiplicative constants, these
are, respectively, the familiar ``singlet" ($\ket{B^0}$) and ``triplet"
$(\ket{B^{1,2,3}})$ Bell states encountered in non-relativistic quantum
information science. The conventional assignment of the four Bell
states\cite{nielsen} is related to the present one by
$\ket{\beta_{11}}=\ket{B^0}, \ket{\beta_{10}} = -i\ket{B^1}, \ket{\beta_{00}} =
\ket{B^2},\ket{\beta_{01}} = i\ket{B^3}$. But we would like to advocate the
convention in Eq.(\ref{e:con}) because it is the particular combination of
$(\tsigma^\mu\sigma^2)_{s s'}a^\dag(\p_1,s; n_1)a^\dag(\p_2,s';n_2)\ket{0}$
which provides us with complete and concise description of the behaviour of
Bell states under {\it arbitrary} Lorentz transformations. The upshot is that
under generic $\Lambda$, these Bell states undergo a rotation among
themselves, and transform as
\begin{equation}\label{e:tyt}
\ket{B^\mu(\p_1,\p_2)}' = U(\Lambda)\ket{B^\mu(\p_1,\p_2)} =
\sqrt{\frac{(\Lambda p_1)^0(\Lambda p_2)^0}{p_1^0p^0_2}}\,
R_\nu\,^\mu(\Lambda,p_1,p_2)\ket{B^\nu({\p_1}_\Lambda, {\p_2}_{\Lambda})},
\end{equation} for which $R_\nu\,^\mu(\Lambda,p_1,p_2) \in SO(4)$ is explicitly listed in
Appendix B.

\subsection{Bell states, and the isomorphism between $[SU(2)\times{SU(2)}]/{Z_2}$ and $SO(4)$}

We shall briefly recap the group isomorphism to establish the notations, and to
explain our line of reasoning. Let us label elements of the two distinct
$SU(2)$ groups by ${\cal U}_{1,2}(\Lambda)$. For our purposes, and following
the analysis of Wigner rotations performed earlier, we should explicitly use
\begin{eqnarray*}
 {\cal U}_1(\Lambda) &=& \exp(i\frac{\bm\sigma}{2}\cdot{\bm\theta}_W(\p_1))  \\
 {\cal U}_2(\Lambda) &=& \exp(i\frac{\bm\sigma}{2}\cdot{\bm\theta}_W(\p_2)),
\end{eqnarray*}
for which $\bm\theta_W(\p_{1,2})$ are the Wigner angles for particles with
momenta $\p_{1,2}$. To set up the group isomorphism, we may consider
\begin{equation}\label{e:xform}
  \mathcal{X}\equiv x_\mu\tsigma^\mu =
  \begin{pmatrix} ix_0+x_3 & x_1-ix_2 \\ x_1+ix_2 & ix_0-x_3 \end{pmatrix}
  =\begin{pmatrix} v & w \\ w^* & -v^* \end{pmatrix} .\end{equation}
  Given ${\cal U}_{1,2}\in SU(2)$, it can be verified that $\mathcal{X}'
\equiv {\cal U}_1 {\mathcal X} {\cal U}_2^{-1}$ is also of the form
\begin{equation}
  \mathcal{X}' = \begin{pmatrix} v' & w' \\ w'^* & -v'^* \end{pmatrix}.
\end{equation}
It follows that we may write $\mathcal{X}' = x'_\mu\tsigma^\mu$. Since $
\det\mathcal{X} = -x_\mu x_\mu=\det\mathcal{X}' =  -x'_\mu x'_\mu$,
 this implies for each $x_\mu$ there is an $R_\mu^{~\nu} \in SO(4)$ such that
$ x'_\mu = R_\mu^{~\nu}x_\nu$. Hence we infer
\begin{equation}
\mathcal{X}' = {\cal U}_1 (x_\mu\tsigma^\mu) {\cal U}_2^{-1} =
R_\mu^{~\nu}x_\nu\tsigma^\mu,\end{equation} yielding the identity
\begin{equation}\label{e:sigmatrans}
   {\cal U}_1\tsigma^\mu{\cal U}_2^{-1} = R_\nu^{~\mu} \tsigma^\nu.
\end{equation}
Returning to one-particle states, the Lorentz transformation for each party of
the entangled pair is (disregarding, for the moment, normalization factors
which are not relevant to the discussion below)
\begin{eqnarray*}
  \ket{\p_1,s,n_1}  &\longmapsto& {\cal U}_1(\Lambda)_{s' s}\ket{{\p_1}_{\Lambda},s',n_1} \\
  \ket{\p_2,s,n_2} &\longmapsto& {\cal U}_2(\Lambda)_{s' s}\ket{{\p_2}_{\Lambda},s',n_2}.
  \end{eqnarray*}
The crucial observation is that the quadruplet ($\mu = 0,1,2,3)$ of states
defined by $(\tsigma^\mu\sigma^2)_{s s'} \ket{\p_1, s, n_1; \p_2, s';n_2}$
transforms as
\begin{eqnarray}
(\tsigma^\mu\sigma^2)_{s s'} \ket{\p_1,s,n_1;\p_2,s',n_2} &\longmapsto&
[{\cal U}_1(\tsigma^\mu\sigma^2){\cal U}_2^T]_{s s'}\ket{{\p_1}_{\Lambda},s,n_1;{\p_2}_{\Lambda},
s',n_2 }   \\
&=& [{\cal U}_1 \tsigma^\mu{\cal U}^{-1}_2\sigma^2]_{s s'}\ket{{\p_1}_{\Lambda},s,n_1;
{\p_2}_{\Lambda},s',n_2 } \\
&=& R_\nu^{~\mu} (\tsigma^\nu\sigma^2)_{s
s'}\ket{{\p_1}_{\Lambda},s,n_1;{\p_2}_{\Lambda}, s',n_2 }.
\end{eqnarray}
In arriving at the last result we have used $\sigma^2 {\cal U}^T_2 ={\cal
U}^{-1}_2\sigma^2$ which follows from $\sigma^2\sigma^i\sigma^2=
-(\sigma^i)^T$; as well as the identity of Eq.(\ref{e:sigmatrans}) in the
final step. As a consequence, Eq.(\ref{e:tyt}) is therefore valid. Furthermore
Eq.(\ref{e:sigmatrans}) also yields the relation
\begin{equation}
  R_\nu^{~\mu}(\Lambda) = \frac{1}{2}\eta_{\nu\alpha}Tr[{\cal U}_1(\Lambda)\tsigma^\mu
                 {\cal U}_2^{-1}(\Lambda) \tsigma^\alpha].
\end{equation}

In passing we mention two special cases:\\
For Lorentz transformations which are pure rotations, $(\X = -\Y = \bm\theta$
as $\bm\phi(\p_{1,2}) =0$); the explicit form of the $SO(4)$ matrix in
Appendix B then yields
\begin{equation*}
  R_\nu^{~\mu} = \begin{pmatrix}
    1      & & {\bm 0}  \\  \\
    \bm{0} & & \cos\theta\delta_{ij} +\epsilon_{ijk}(\sin\theta)\hat \theta^k +
    (1-\cos\theta)\hat \theta^i
    \hat \theta^j
  \end{pmatrix};
\end{equation*}
which means that, as expected, the three Bell states $\ket{B^{1,2,3}(\p_1,
\p_2)}$ form an $SO(3)$ {\it rotation triplet} while $\ket{B^0(\p_1, \p_2)}$
is a {\it singlet}. For an equal-mass entangled pair in the
center-of-momentum(COM) frame ($\p_1 + \p_2 = 0$) and pure boost in the
perpendicular direction (${\bm\tau}\cdot\p_{1,2} = 0$), Eqs. (A9)-(A13) of
Appendix A reveal that the resultant Wigner rotations are related by
$\bm\phi(\bm\tau,\p_1) = -\bm\phi({\bm\tau},\p_2)\equiv \bm\phi$. Thus with
($\X = \Y = \bm\phi$) the result is
\begin{equation*}
  R_\nu^{~\mu} =
  \begin{pmatrix}
    \cos\phi             & & (\sin\phi)\hat{\phi^j} \\  \\
    -(\sin\phi)\hat{\phi^i} & & \delta_{ij}+(\cos\phi - 1)\hat \phi^i\hat \phi^j
  \end{pmatrix}.
\end{equation*}
In general, rotational singlet and triplet states do not belong to invariant
subspaces when boost transformations are also included. Special cases of the
behaviour of Bell states for massive particles under Lorentz transformations
have also been calculated in Refs.\cite{milburn,ahn,terashima}. As we have
shown the precise and complete behaviour under {\it arbitrary} Lorentz
transformations can be succinctly stated, as in Eq.({\ref{e:tyt}}).

\subsection{Two-particle states as superpositions of Bell states}

In the previous sections we discussed Bell states with infinitely sharp
momenta, but it is possible to generalize the discussion to generic
superpositions of Bell states
\begin{equation}
\ket{\Psi} = \int\,d^3\p_1 \,\,\int\,d^3\p_2\, C_\mu(\p_1,n_1;\p_2,n_2
)\ket{B^\mu(\p_1, n_1;\p_2, n_2)}.
\end{equation}
Moreover, it is actually possible to think of any two-particle state, {\it
entangled or otherwise}, of the form
\begin{equation}
\ket{\Psi} = \int\,d^3\p_1 \,\,\int\,d^3\p_2\, f(\p_1,s_1,n_1
;\p_2,s_2,n_2)a^\dag(\p_1,s_1,n_1)a^\dag(\p_2,s_2,n_2)\ket{0}
\end{equation}
in terms of Bell states. Clearly a relation between the coefficients given by
\begin{equation}
f(\p_1,s_1,n_1;\p_2,s_2,n_2) =
C_\mu(\p_1,n_1;\p_2,n_2)(\tsigma^\mu\sigma^2)_{s_1 s_2}
\end{equation}
works.
The relation is invertible as
\begin{equation}
C_\mu(\p_1,n_1;\p_2,n_2) =
\frac{1}{2}\eta_{\mu\nu}f(\p_1,s_1,n_1;\p_2,s_2,n_2)(\sigma^2\tsigma^\nu)_{s_2
s_1}
\end{equation}
Thus $C_\mu(\p_1,n_1;\p_2,n_2)$ can be written down given
$f(\p_1,s_1,n_1;\p_2,s_2,n_2)$, and vice versa.
Note however that in quantum field theory all states transform unitarily
$(\ket{\Psi}' = U(\Lambda)\ket{\Psi})$ under Lorentz transformations, no
matter how complicated the superposition is. The coefficients
$f(\p_1,s_1,n_1;\p_2,s_2,n_2)$ and $C_\mu(\p_1,n_1;\p_2,n_2)$ are not
operator-valued, and commute with $U(\Lambda)$.

\section{Reduced density matrices, identical particles, and Lorentz-invariance of von Neumann entropy}

We next introduce reduced density matrices and their properties. While it is
possible to generalize, we shall choose to concentrate on systems with
identical spin $\frac{1}{2}$ massive fermions e.g. electrons.

Given an ${\bbN}$-particle system of identical particles with density matrix
$\rho$ (which is not restricted to a pure state density matrix, but may also
correspond to a mixed configuration $\Tr{\rho^2} \neq \Tr\rho$), the
$m$-particle ($m < \bbN$) reduced density matrices can be defined as
\begin{equation}\label{e:yy}
  \rho_m \equiv \frac{1}{(m!)^2}\ket{i_1 i_2...i_m}\Tr\{a_{i_1}a_{i_2}...a_{i_m}
  \rho\,a^\dag_{j_m}a^\dag_{j_{m-1}}...a^\dag_{j_1} \}\bra{j_1j_{2}...j_m}.
\end{equation}
This is equivalent to Yang's definition\cite{yang}. Note that $\rho_m$ is an
m-particle operator; and we have simplified all the quantum numbers of the
creation operator $a^\dag_{i_k}$ to the label $i_k$. It can then be worked out
that Eq.(\ref{e:yy}) implies
\begin{eqnarray}\label{e:vev}
  \frac{\bra{i_1...i_m}\rho_m\ket{j_1...j_m}}{m!(\Tr\rho_m)} &=&
    \frac{\bra{i_{1}...i_m{k_{m+1}}...k_n}\rho_n\ket{j_1...j_mk_{m+1}...k_{n}}}{n!(\Tr\rho_n)} \qquad
    \forall ~ m<n ~,~ 1 < n \leq {\bbN},\\
\text{with} \qquad\frac{\rho_{\bbN}}{\Tr\rho_{\bbN}} &=&
\frac{\rho}{\Tr\rho}\qquad .
\end{eqnarray}
Thus these reduced density matrices are defined by the partial traces of higher
particle number density operators.

It is worth emphasizing in quantum field theory Lorentz transformations are
implemented {\it unitarily} on {\it physical states}. Under any Lorentz
transformation $\Lambda$, all creation and annihilation operators transform as
$U(\Lambda){a^\dag}_i {U^\dag}(\Lambda)$ and $U(\Lambda){a}_i
{U^\dag}(\Lambda)$. It follows that (we also assume a Lorentz invariant vacuum
$U(\Lambda)\ket{0}=\ket{0}$) all states obtained through the action of creation
and annihilation operators on the vacuum must transform unitarily as
$\ket{\Psi}' = U(\Lambda)\ket{\Psi}$ and
$\bra{\Psi}'=\bra{\Psi}U^\dag(\Lambda)$. As a consequence, under Lorentz
transformations all reduced density matrices defined above also transform
unitarily as $\rho' = U(\Lambda)\rho U^\dag(\Lambda)$.

The von Neumann entropy of a density matrix is
\begin{equation}
S \equiv -\Tr{(\rho\ln\rho)} = -\sum^N_{n=1} \lambda_n\ln\lambda_n.
\end{equation}
with $\lambda_n$ being the eigenvalues of $\rho$ (we assume normalization of
$\Tr\rho = 1$ has been carried out in the definition of the von Neumann
entropy). Since $\rho$ is Hermitian it can be diagonalized, and we may write
$\rho = V[diag(\lambda_1,..., \lambda_{N})]V^{\dagger}$. It follows that the
eigenvalues are {\it invariant} under unitary transformations (and, in
particular, Lorentz transformations) since $\rho' = U(\Lambda)\rho
U^\dag(\Lambda) = U(\Lambda)V[diag(\lambda_1,...,
\lambda_{N})][{U(\Lambda)V}]^{\dagger}$ obviously has the same eigenvalues as
$\rho$. Thus the von Neumann entropy is {\it Lorentz invariant}. Moreover the
von Neumann entropy of {\it all} reduced density matrices defined by $S_m
\equiv -\Tr{(\rho_m\ln\rho_m)}$ with $(m=1,...,{\bbN})$ are also invariant for
the same reasons. A physical system defined by $\rho$ can thus be parametrized
by a set of Lorentz invariant measures $\{S_1,..., S_m,..., S_{\bbN}=S\}$. We
shall proceed to show that $S_m$ can be useful measures of entanglement
shortly.

\subsection{Reduced density matrix, Bell states, and Lorentz-invariant entanglement: a worked example}

Consider a system of two identical fermions, and for the first part of the
illustration let us follow the discussion of Ref.\cite{paskauskas}. A
two-particle state of identical fermions may be written as
\begin{equation}
  \ket\Psi = C_{ij}a^\dag_i a^\dag_j\ket{0}  \qquad i,j=1,2,\cdots,{\cal N}.
\end{equation}
$C_{ij}$ is anti-symmetric and can be set into block diagonal form\cite{yang}
via
\begin{equation}\label{e:k}
  {\cal U}C{\cal U}^T = \bigoplus_{i=1}^{{\cal N}_f} \begin{pmatrix} 0 & c_i \\ -c_i & 0 \\
  \end{pmatrix}
   \overbrace{\bigoplus(0)}^\text{for ${\cal N}$ odd};
\end{equation}
with ${\cal N}_f \equiv ({\cal N}/2)$ for even $\cal N$, and for odd ${\cal
N}$, ${\cal N}_f \equiv ({\cal N}-1)/2$; and ${\cal U}$ is a unitary matrix.
Considering a redefinition with $a'^\dag_i = {\cal U}^*_{ij}a^\dag_j$, we may
also rewrite
\begin{equation}
  \ket{\Psi} = 2\sum_{i=1}^{{\cal N}_f} c_i a'^\dag_i a'^\dag_{i+1} \ket{0}.
\end{equation}
  This is the analog of ``Schmidt
decomposition" for identical fermion systems\cite{schliemann}. The total system
has density matrix $\rho = \ket{\Psi}\bra{\Psi}$ and entropy $S=0$. Following
the prescription for density matrix reduction discussed earlier, the
one-particle reduced density matrix is then
\begin{equation}
{\bra i}\rho_1{\ket j} = \Tr{\{a_i\rho\, a^\dag_j}\} = 4(CC^\dag)_{ij}.
\end{equation}
In terms of (${\cal U}C{\cal U}^T$) combination of Eq.(\ref{e:k}), we note that
$\rho_1 = 4{\cal U}^\dag[({\cal U}C{\cal U}^T)({\cal U}C{\cal U}^T)^\dag]({\cal
U})$; thus
\begin{equation}\label{e:gg}
 {{\cal U}\rho_1{\cal U}^\dag} = 4\bigoplus_{i=1}^{{\cal N}_f}
 \begin{pmatrix} {\abs{c_i}}^2 &0\\ 0& {\abs{c_i}}^2 \\ \end{pmatrix}
   \overbrace{\bigoplus(0)}^\text{for ${\cal N}$ odd}.
\end{equation}
The normalization condition is
\begin{equation}\label{e:h}
{\langle \Psi|\Psi\rangle} = 4\sum_{i=1}^{{\cal N}_f}{\abs{c_i}}^2=1
\qquad\Rightarrow \Tr\rho_1 = 8\sum_{i=1}^{{\cal N}_f}{\abs{c_i}}^2 = 2;
\end{equation}
and the von Neumann entropy of the {\it reduced} density matrix is therefore
\begin{equation}
  S_1 = -\Tr(\frac{\rho_1}{\Tr\rho_1}\ln\frac{\rho_1}{\Tr\rho_1})
  = -4\sum_{i=1}^{{\cal N}_f}{\abs{c_i}}^2\ln(2{\abs{c_i}}^2);
\end{equation}
which is bounded by $\ln 2 \leq S_1 \leq \ln(2{\cal N}_f)$. The upper limit is
obtained by maximizing $S_1$ subject to the constraint (\ref{e:h}); while the
lower bound occurs when there is only a single non-vanishing anti-symmetric
block in Eq.(\ref{e:k})\cite{paskauskas,schliemann}. It may appear
disconcerting that this lower bound for fermions is $\ln 2$, rather than zero,
whereas the analogous result for a system of two bosons is $0\leq S_1\leq\ln
{\cal N}$\cite{paskauskas}. However, the lowest entropy is associated with
what we would call an ``unentangled" system; while any of the ``entangled"
Bell state discussed earlier will be shown to have $S_1 = 2\ln2$ instead. To
see this, let us use, for instance, an unentangled two-fermion state,
\begin{eqnarray}
\ket\Psi &=&\frac{1}{\sqrt{2}}[\ket{B^0(\p_1,\p_2)}+i\ket{B^3(\p_1,\p_2)}]\\
&=& {a^\dag(\p_1,+)a^\dag(\p_2,-)}\ket{0}\\
&\equiv& C_{ij}a^\dag_i a^\dag_j\ket{0}; \end{eqnarray} yielding the
non-vanishing part of the C-matrix as \begin{math}
  C= \bordermatrix{&(\p_1,+)&\p_2,-)\cr
          (\p_1,+) & 0  & \frac{1}{2} \cr
          (\p_2,-) & -\frac{1}{2} & 0 \cr}.
\end{math}
Following earlier derivations, we obtain the normalized one-particle reduced
density matrix as
\begin{equation}
{\rho_1} =\begin{pmatrix} &\half & 0 \cr & 0 & \half \end{pmatrix};
\end{equation}
giving $S_1 = -2(\frac{1}{2})\ln\frac{1}{2} = \ln 2$, which is the {\it
minimum} von Neumann entropy of the reduced 1-particle density matrix for two
identical fermions. We chose to compute $C$ explicitly to illustrate the
consistency of the approach, but it is not necessary to go through this step.
Given $\rho$, computation of $\rho_1$ can also be done directly through
Eq.(\ref{e:vev}).

Consider instead the Bell state
\begin{eqnarray}
\ket\Psi &=&\ket{B^0(\p_1,\p_2)}\\
&=&\frac{1}{\sqrt{2}}[{a^\dag(\p_1,+)a^\dag(\p_2,-)}-{a^\dag(\p_1,-)a^\dag(\p_2,+)}]\ket{0}\\
&\equiv& C_{ij}a^\dag_i a^\dag_j\ket{0}; \end{eqnarray} now yielding the
non-vanishing part of the $C$-matrix as
\newline
\begin{math}
C= \bordermatrix{&(\p_1,+)&(\p_2,-)&(\p_2,+)&(\p_1,-)\cr
          (\p_1,+)& 0 & \frac{1}{2\sqrt{2}}  & 0  & 0 \cr
          (\p_2,-)& -\frac{1}{2\sqrt{2}} & 0  & 0  & 0  \cr
          (\p_2,+) & 0 & 0  & 0  & \frac{1}{2\sqrt{2}}  \cr
          (\p_1,-) & 0  & 0  & -\frac{1}{2\sqrt{2}}  & 0 \cr},
\end{math} for which the normalized reduced density matrix is
\begin{equation}
{\rho_1} =\begin{pmatrix} \frac{1}{4} & 0 & 0\\
0 & \frac{1}{4} & 0 & 0\\
0 & 0 & \frac{1}{4} & 0\\
0 & 0 & 0 & \frac{1}{4}
\end{pmatrix}.
\end{equation}
The corresponding entropy is now $S_1 = -4(\frac{1}{4}\ln \frac{1}{4}) =
2\ln2$ instead. This is a von Neumann entanglement entropy (of the reduced
density matrix) which is $\ln2$ units {\it higher} than the lowest value thus
justifying that the Bell state $\ket{B^0(\p_1,\p_2)}$ is an entangled state.
Similarly all the Bell states $\ket{B^\mu(\p_1,\p_2)}$ discussed previously
have von Neumann entanglement entropy of value $2\ln2$. Moreover, as
explained, for any physical system the von Neumann entropy of any of the
reduced density matrices in quantum field theory will be {\it Lorentz
invariant}.

\subsection{Generalized zeta functions and von Neumann Entropy}

Zeta function regularizations have been employed in quantum field theory as
gauge and Lorentz-invariant methods for taming and isolating
divergences\cite{hawking,critchley}. Here we briefly touch upon its relation
to the von Neumann entropy.

The generalized zeta function of an operator can be defined to be
\begin{equation}
  \zeta_{\hat O}(s) \equiv \sum_n\frac{1}{o_n^s} \end{equation}
where $o_n$ are eigenvalues of the operator $\hat O$. There is an interesting
relation between generalized zeta function and von Neumann entropy. Recall that
a density matrix $\rho$ (or, for this matter, $\rho_m$, a reduced density
matrix) has von Neumann entropy
\begin{equation}\label{e:415}
S = -Tr({\rho \ln \rho}) = -\sum_n \lambda_n \ln\lambda_n
\end{equation}
with $\lambda_n$ being the eigenvalues of $\rho$. Note that in the sum, zero
eigenvalues do not pose a problem (by L'Hospital rule $\lambda_n\ln\lambda_n$
has no contribution for zero eigenvalues). However in quantum field theory,
without regularization, the entropy can suffer from divergences. For instance,
$\rho = {e^{-\beta H}}/{\Tr({\exp^{-\beta H}})}$ leads to the thermodynamic
relation $S = \beta\langle H \rangle$, and the expectation value is in general
divergent in quantum field theory.

If we adopt ${\hat O} =\rho$, then $\zeta_{\rho}(s) =
\sum_n\frac{1}{\lambda_n^s} = \sum_n \exp(-s\ln\lambda_n)$. We note that
\begin{eqnarray}
{\frac{d\zeta_{\rho}}{ds}}\bigg\vert_{s=-1} &=& -\lim_{s=-1} \sum_n
\exp(-s\ln\lambda_n)\ln\lambda_n\\
&=&-\sum_n \lambda_n \ln\lambda_n.
\end{eqnarray}
Thus it is possible to define $S(s) \equiv {\frac{d\zeta_{\rho}}{ds}}(s)$, and
analytically continue from values for which it is defined to $s=-1$. However,
depending on the form of the operator $\rho$, it may still be that $s=-1$ is a
pole of $\zeta_\rho(s)$.

An alternative then is to note the sum for $S$ in Eq.(\ref{e:415}) does not
include zero eigenvalues. Therefore we may substitute $\rho$ with its
sub-matrix $\rho'$ which contains no zero eigenvalues. Its inverse,
${\rho'}^{-1}$, exists; and we may construct its generalized zeta function as
\begin{equation}
  \zeta_{[\rho']^{-1}}(s) = \sum_{n} (\lambda_n)^s.\end{equation}
Since $\rho$ is a density matrix, its eigenvalues satisfy $0 \leq \lambda_n
\leq 1 \,\,\forall n$; so for positive $s$, the sum is bounded above by the
dimension of $\rho$, and therefore converges for any finite-dimensional $\rho$.
Furthermore
\begin{eqnarray}
S(s)\bigg\vert_{s=1} \equiv
-{\frac{d\zeta_{{\rho'}^{-1}}(s)}{ds}}\bigg\vert_{s=1} &=& -\lim_{s=1} \sum_n
\exp(s\ln\lambda_n)\ln\lambda_n\\
&=&-\sum_n \lambda_n \ln\lambda_n,
\end{eqnarray}
and for infinite-dimensional $\rho$ we may also define the von Neumann entropy
$S$ by analytic continuation of the zeta function $\zeta_{{\rho'}^{-1}}(s)$ and
$S(s)$ to $s=1$. A further generalization is
\begin{equation}
S \equiv \frac{1}{\alpha}\frac{d\zeta_{\rho^\alpha}(s)}{ds}\bigg\vert_{s =
{-1}/{\alpha}},
\end{equation}
with $\rho'$ substituting for $\rho$ when $\alpha < 0$.

\section{Further comments and conclusions}

Our work links the Wigner rotations of spins to the behaviour of Bell States
under arbitrary Lorentz transformations. We discussed reduced density matrices
for identical particle systems, established the Lorentz invariance of their von
Neumann entropies, and suggested an invariant regularization through
generalized zeta function. In addition, we worked out the explicit expressions
of the infinite-dimensional Hermitian generators in the momentum
representation. We hope that the results presented here will help to place
Relativistic Quantum Information Science on the surer foundations of quantum
field theory which is fully compatible with Lorentz symmetry and causality. It
is worth emphasizing that we considered the {\it full} reduced density matrix
with all the degrees of freedom, including the momentum with the reduction
prescription of Section V when we go from $\bbN$ particles to $m < {\bbN}$
particles.  This is different from the reduced density matrix used, for
instance in Ref.\cite{peres}, in which ``reduction" to the spin degree of
freedom is applied even to a single particle. The von Neumann entropy is
invariant in our case because Lorentz symmetry is unitarily implemented in
quantum field theory; each creation operator transforms unitarily under the
Lorentz group, hence the prescribed reduction of density matrices in Section V
will result in Lorentz-invariant von Neumann entropy for whatever resultant
reduced density matrix we have. In contradistinction, Lorentz symmetry is not
implemented unitarily in non-relativistic treatments of ``wavefunctions".

It is known through earlier efforts by others that for a system of two
particles in a total pure state, the von Neumann entropy of the reduced
density matrix is a good measure of the entanglement. We demonstrated that
such a characterization is in fact Lorentz invariant. Although a system of $n$
particles can be similarly parametrized by the Lorentz-invariant von Neumann
entropy each of the reduced density matrices, the characterization of
entanglement in terms of these $n$ numbers, and especially in the case when
the total system is a mixed configuration rather than a pure state, is still
not completely clear and merits further studies.



\begin{acknowledgments}
The research for this work has been supported by funds from the National
Science Council of Taiwan under Grants No. NSC91-2112-M006-018 and
NSC92-2120-M-006-004. We would like to thank all members of the QIS group,
especially C.C. Chen, C. Cheng, and W. M. Zhang for helpful feedback. C.S. is
also grateful to Kuldip Singh for beneficial discussions on topics related to
the present article, and especially to Daniel Terno for very useful
correspondence on finite Wigner rotations.

\end{acknowledgments}

\appendix
\section{Computation of the complete Wigner angle}
\newcommand*{\mn}{\protect{\mu\nu}}
\newcommand*{\ab}{\protect{\alpha\beta}}
\newcommand*{\fiq}{\ensuremath{\frac{i}{4}}}
\newcommand*{\coef}{\left(\frac{p^0}{m}-1\right)}

We shall first calculate the infinitesimal Wigner angle by Taylor expansion as
\begin{eqnarray}  \label{e:1}
  W(\Lambda, p) &=& W(\Lambda,p)\bigg\vert_{\substack{\omega=0 \\ (\Lambda=1)}} +
\frac{\omega_{\alpha\beta}}{2}\frac{dW}{d\omega_{\alpha\beta}}\bigg\vert_{\substack{\omega=0 \\
(\Lambda=1)}} + \cdots\\
&=& I + \frac{\omega_{\alpha\beta}}{2}[\frac{dL^{-1}(\Lambda p)}
{d\omega_{\alpha\beta}}]\bigg\vert_{\substack{\omega=0}}L(p)+
\frac{\omega_{\alpha\beta}}{2}L^{-1}(p)\frac{d\Lambda(\omega)}{d\omega_{\alpha\beta}}
\bigg\vert_{\omega=0}L(p) + \cdots\label{e:wiglast}\\
 &=&  \I -
{\frac{\omega_\ab}{2} [L^{-1}(\Lambda p)\frac{dL(\Lambda
  p)}{d\omega_\ab}}]\bigg\vert_{\Lambda=1} +
  \frac{i\omega_{\alpha\beta}}{2}L^{-1}(p)M^{\alpha\beta}L(p).
\end{eqnarray}
to first order in $\omega$; with $\,\Lambda(\omega) = \I+\ihalf\omega_{\mu\nu}
M^{\mu\nu}\,{\rm and}\,
  \Lambda^\mu\,_\nu  = p^\mu+\ihalf(\omega_\ab
  M^\ab)^\mu_{~\nu}p^\nu =\delta^\mu_{~\nu}+\omega^\mu_{~\nu}$.
Straightforward and careful calculations using the explicit matrix elements,
\begin{eqnarray} \label{e:repre}
  (K^i)_{ab} &=& i(\delta^i_{a}\delta_{0b}+\delta_{0a}\delta^i_{b})  \\
  (J^i)_{ab} &=& -i\epsilon_{iab},
\end{eqnarray}
of the generators $K^i = M^{0i}$ and $ J^i = \frac{1}{2}\epsilon_{ijk}M^{jk}$,
and the expression of the Lorentz boost of Eqs.(2.4-2.6) yield the complete
infinitesimal Wigner rotation as
\begin{equation}\label{e:wig}
W(\Lambda,p) = \I
+\ihalf\left[\omega_{ij}-\frac{1}{p^0+m}(p_i\omega_{j0}-p_j\omega_{i0})\right]
M^{ij}
\end{equation}
with $p^0 \equiv \sqrt{\np^2 + m^2}$.

Eq.(\ref{e:wiggen}) implies Wigner transformations satisfy the multiplication
rule
\begin{equation} W(\Lambda_2\Lambda_1, p) = W(\Lambda_2,\Lambda_1 p)\cdot
W(\Lambda_1,p).\label{e:multi} \end{equation} To compute the Wigner rotations
for general Lorentz transformations with 6 independent parameters, we may
consider the decomposition $\Lambda = R({\bm\psi})\cdot L({\bm\alpha})$. The
multiplication rule leads to
\begin{equation} W(\Lambda, p) = W(R(\bm{\psi}),L({\bm\alpha})p)\cdot
W(L({\bm\alpha}),p).\label{e:mul} \end{equation} Regardless of the momenta,
Wigner angles are degenerate with ordinary rotation angles when the boost
parameters are zero; thus the first factor is just $\exp(i{\bm\psi}\cdot{\bm J}
)$, while the second factor is the Wigner rotation for an arbitrary pure
boost. Exploiting the homomorphism between $SL(2,C)$ and $SO(3,1)$, the Wigner
angle of this remaining factor has been successfully computed by
Halpern\cite{halpern} by direct multiplication of $2\times2$ $SL(2,C)$
matrices representing the Lorentz transformations in Eq.(2.4-2.6). We may
express the result of Halpern as
\begin{equation}
W(L({\bm\tau}),p) \equiv \exp(i{\bm\phi(\bm\tau)}\cdot{\bm J}),
\end{equation}
with \begin{eqnarray} \cos\phi&=&\frac{[\cosh\tau + \cosh\chi +
\sinh\tau\sinh\chi(\hat{\bm\tau}\cdot{\hat\p}) + (\cosh\tau-1)(\cosh\chi -1)
(\hat{\bm\tau}\cdot{\hat\p})^2]}{[1 + \cosh\tau\cosh\chi +
\sinh\tau\sinh\chi(\hat{\bm\tau}\cdot{\hat\p})]}
\\
&=&\frac{[m\cosh\tau + p^0 + \sinh\tau(\hat{\bm\tau}\cdot{\p}) +
(\cosh\tau-1)(p^0 -m) (\hat{\bm\tau}\cdot{\hat\p})^2]}{[m + p^0\cosh\tau +
\sinh\tau(\hat{\bm\tau}\cdot{\p})]},
\\
 (\sin\phi){\hat{\bm\phi}} &=& \frac{[\sinh\tau\sinh\chi +
(\cosh\tau-1)(\cosh\chi-1)(\hat{\bm\tau}\cdot{\hat\p})]}{[1 +
\cosh\tau\cosh\chi + \sinh\tau\sinh\chi(\hat{\bm\tau}\cdot{\hat\p})]}
(\hat{\bm\tau}\times{\hat\p})\\
&=&\frac{[|\p|\sinh\tau +(p^0 -m)(\cosh\tau
-1)(\hat{\bm\tau}\cdot{\hat\p})]}{[m + p^0\cosh\tau +
p\sinh\tau(\hat{\bm\tau}\cdot{\hat\p})]}(\hat{\bm\tau}\times{\hat\p});
\end{eqnarray}
and the rapidity $\chi$ is related to $\p$ by $\sinh\chi = \frac{|\p|}{m},
\cosh\chi = \frac{p^0}{m}$. It is easy to confirm the infinitesimal limit is
indeed
\begin{equation}
{\bm \phi}(\bm\tau) \longrightarrow {\bm \phi}_1 = -\frac{\p \times
\vtau}{p^0+m}.
\end{equation}
The complete expression of the Wigner rotation of (A12) is therefore
\begin{equation}
W(\Lambda, p) = \exp(i{\bm\theta}_W\cdot{\bm J}) = \exp(i{\bm\psi}\cdot{\bm J}
)\cdot\exp(i{\bm\phi}({\bm\alpha})\cdot{\bm J});
\end{equation}
which yields the explicit relations
\begin{eqnarray}
\cos(\frac{\theta_W}{2}) &=& (\cos\frac{\psi}{2})(\cos\frac{\phi}{2}) -
(\sin\frac{\psi}{2})(\sin\frac{\phi}{2})(\hat{\bm\psi}\cdot\hat{\bm\phi}),\\
\sin(\frac{\theta_W}{2})\hat{\bm\theta}_W
&=&(\cos\frac{\psi}{2})(\sin\frac{\phi}{2})\hat{\bm\phi}
+(\sin\frac{\theta}{2})(\cos\frac{\phi}{2})\hat{\bm\psi} +
(\sin\frac{\psi}{2})(\sin\frac{\phi}{2})(\hat{\bm\phi}\times\hat{\bm\psi}).
\end{eqnarray}
\section{Explicit form of the rotation matrix $R_\nu^{~\mu}(\Lambda,p_1,p_2)$}

The rotation matrix among the four Bell states has been shown to be
\begin{eqnarray}
  R_\nu^{~\mu} &=&\frac{1}{2}\eta_{\nu\alpha}\Tr[{\cal U}_1(\Lambda) \tsigma^\mu
                 {\cal U}^{-1}_2(\Lambda) \tsigma^\alpha] \\
&=&\frac{1}{2}\eta_{\nu\alpha}\Tr[\exp(i\X\cdot\frac{\bm\sigma}{2}){\tsigma^\mu}
                  \exp(i\Y\cdot\frac{\bm\sigma}{2}){\tsigma^\alpha}];
\end{eqnarray}
with the definitions \begin{eqnarray}
 \X &\equiv&  {\bm\theta}_W(\p_1), \\
 \Y &\equiv& -{\bm\theta}_W(\p_2).\\
\end{eqnarray}
Straightforward computations yield the explicit matrix elements
\begin{eqnarray*}
  R_0^{~0} &=& (\cos\hx)(\cos\hy) - \hat X\cdotp\hat Y (\sin\hx)(\sin\hy) \\
  R_i^{~0} &=& -R^0_{~i} = - (\cos\hx)(\sin\hy)\hat Y_i - (\sin\hx)(\cos\hy)\hat X_i
    + (\sin\hx)(\sin\hy)\epsilon_{ijk}\hat X_j\hat Y_k \\
  R_i^{~j} &=& (\cos\hx)(\cos\hy)\delta_{ij} - (\cos\hx)(\sin\hy)\epsilon_{ijm}\hat Y_m
    + (\sin\hx)(\cos\hy)\epsilon_{ijm}\hat X_m + (\sin\hx)(\sin\hy)[\hat X\cdotp\hat Y\delta_{ij}
    -\hat X_i\hat Y_j-\hat X_j\hat Y_i].
\end{eqnarray*}



\end{document}